\documentclass[sigconf]{acmart}

\newcommand{\slfrac}[2]{\left.#1\middle/#2\right.}
\DeclareMathOperator*{\argmin}{arg\,min}
\DeclareMathOperator*{\argmax}{arg\,max}

\setcopyright{none} 
\acmConference[Anonymous Submission to ACM CCS 2017]{ACM Conference on Computer and Communications Security}{Due 19 May 2017}{Dallas, Texas}
\acmYear{2017}

\usepackage{graphicx}
\usepackage{amsmath}
\usepackage{textcomp}
\usepackage{subfig}
\usepackage{url}
\usepackage{algorithm}
\usepackage[noend]{algpseudocode}
\usepackage[toc,page]{appendix}
\usepackage{todonotes}

\begin{document}
\title{Mitigation of Adversarial Attacks through \\ Embedded Feature Selection} 

\author{Ziyi Bao}
\affiliation{Imperial College London}
\email{ziyi.bao14@imperial.ac.uk}

\author{Luis Mu\~{n}oz-Gonz\'{a}lez} 
\affiliation{Imperial College London}
\email{l.munoz@imperial.ac.uk}

\author{Emil C. Lupu}
\affiliation{Imperial College London}
\email{e.c.lupu@imperial.ac.uk}

\begin{abstract}
Machine learning has become one of the main components for task automation in many application domains. Despite the advancements and impressive achievements of machine learning, it has been shown that learning algorithms can be compromised by attackers both at training and test time. Machine learning systems are especially vulnerable to adversarial examples where small perturbations added to the original data points can produce incorrect or unexpected outputs in the learning algorithms at test time. Mitigation of these attacks is hard as adversarial examples are difficult to detect. Existing related work states that the security of machine learning systems against adversarial examples can be weakened when feature selection is applied to reduce the systems' complexity. In this paper, we empirically disprove this idea, showing that the relative distortion that the attacker has to introduce to succeed in the attack is greater when the target is using a reduced set of features. We also show that the minimal adversarial examples differ statistically more strongly from genuine examples with a lower number of features. However, reducing the feature count can negatively impact the system's performance. We illustrate the trade-off between security and accuracy with specific examples. We propose a design methodology to evaluate the security of machine learning classifiers with embedded feature selection against adversarial examples crafted using different attack strategies.
\end{abstract}

\keywords{Adversarial machine learning, adversarial examples, feature selection.} 

\maketitle

\section{Introduction} \label{sec:1}
Machine learning has proven to be a powerful tool for the automation of many tasks, even outperforming humans in some instances. It offers important benefits in terms of new functionality, personalisation, and optimisation of resources. Given the huge amount of data available from diverse sources, machine learning has become the main component of many systems. Despite the rapid development of new techniques, it has also been shown that machine learning systems are vulnerable. Attackers can perform poisoning attacks by injecting malicious examples into the training dataset \cite{nelsonSpam,biggioSVM,mei,luisPoisoning,jagielski}, subverting the learning process and thus degrading the performance of the system in either a targeted or an indiscriminate manner. Attackers can also exploit the system's weaknesses and blind spots at test time by crafting malicious examples that produce intentional errors and evade detection \cite{huang,biggioEvasion,papernotLimitations,carliniTowards}. Some of these attacks have already been reported in the wild in different applications, such as autonomous bots and malware detection, among others. These vulnerabilities can hinder the adoption of machine learning and have fostered a growing research community in adversarial machine learning \cite{huang,mcdaniel,papernotTowards,luisSecret}, which aims to understand the mechanisms that can allow attackers to compromise or evade the learning algorithms, propose mechanisms to mitigate the effect of these vulnerabilities, and provide new design methodologies for more secure systems.

It has been shown that machine learning models are often vulnerable to \emph{adversarial examples}, i.e. samples that are slightly modified by an adversary to produce intentional errors in the system at test time \cite{biggioEvasion,szegedy,goodfellow,papernotLimitations}. Szegedy et al. \cite{szegedy} have shown show the vulnerabilities of neural networks against adversarial examples in computer vision problems. They have also shown that detecting such attacks is challenging as they usually require adding only small perturbations to the original image, which makes adversarial examples undetectable even to the human eye. This problem has recently led to an increasing number of contributions, especially in the context of deep networks, as for example in \cite{goodfellow,papernotLimitations,carliniTowards,papernotDistillation,tramer2018,madry,kurakinScale,moosavi2016,kurakinPhysical,moosavi2017}.

Defensive techniques attempting to mitigate this problem have been proposed in the research literature. We can broadly classify these defences into two groups: On one side, some techniques aim to classify adversarial examples \cite{gu2014,jin2015,shaham2015,papernotDistillation,zheng2016}. Other defences try to detect them instead \cite{bhagoji,feinman2017,grosse2017,metzen2017,hendrycks2016,li2017}. However, Carlini \& Wagner \cite{carliniBypassing} showed that most of these defences can be defeated if the attacker crafts adversarial examples targeting the specific defence, although some techniques indeed help to mitigate the problem, such as the adversarial training techniques in \cite{madry,tramer2018}.

Adversarial examples are produced in regions where there is a mismatch between the learned model and the ideal model we would obtain if we could completely characterise the underlying data distribution (i.e. if we had an infinite number of training points). There are trivial cases where the attacker can leverage regions not supported by the training data points, where the attacks can be easily detected through appropriate anomaly detection. However, adversarial examples are produced in regions where training data does not occur naturally, or only with very low probability, but are sufficiently close to the genuine data points such that detection becomes extremely challenging, especially in high-dimensional datasets. Goodfellow et al. \cite{goodfellow} claim that, in the case of neural networks, there is an \emph{excess of linearity} in the behaviour of the networks. However, attack \emph{transferability} \cite{szegedy,papernotTransferability} suggest that adversarial regions are shared across different machine learning models, enabling black-box attacks by using substitute models \cite{papernotBlackBox}, which indicates that different learning algorithms suffer from the same limitations. 

Dimensionality reduction through feature selection and extraction have also been explored in the research literature with contradictory conclusions. Bhagoji et al. \cite{bhagoji} showed that linear dimensionality reduction with Principal Component Analysis (PCA) and \emph{anti-whitening} helps to enhance the robustness of machine learning systems against evasion attacks. The other side \cite{biggioSecurity,zhangAdversarial,wangSparse} claims that applying dimensionality reduction through feature selection yields models that are less secure against evasion attacks compared to models that use the full feature set. In \cite{biggioSecurity,zhangAdversarial} filter and wrapper feature selection methods are proposed, which often produce less optimal model than embedded feature selection, such as Lasso. Wang et al. \cite{wangSparse} analyse the security properties of Lasso, but the attacker model is unrealistic under white-box settings. Although in these three works experimental evidence is shown to support the validity of their hypotheses, the proposed experimental methodology relies on metrics that compare the overall perturbation that the attacker needs to introduce to succeed. However, as models with different numbers of features are compared, this does not provide a fair comparison on the perturbation relative to the number of features used. 

In this paper, we show that in linear learning algorithms, embedded feature selection increases the systems' robustness  against adversarial examples, considering minimal white box attacks. We propose a methodology to compare the security of learning algorithms against different evasion attacks for models with different feature counts. We also show that there is a natural trade-off between accuracy and security as a function of the selected number of features. Thus, the contributions of the paper are the following:
\begin{itemize}
\item We propose a methodology to compare the security of machine learning classifiers with different numbers of features against different forms of attacks, according to the norm used to measure the perturbation introduced in the adversarial examples. We propose to use a normalised average minimum distance to measure the relative minimum distortion the required for a successful attack.
\item We propose an aggregated metric to assess the security of the learning algorithms against adversarial examples, considering different types of attacks. 
\item We empirically show that embedded feature selection enhances the robustness of linear machine learning classifiers against adversarial attacks. We characterise the tradeoff between accuracy and security in these models and show that, in some cases, by sacrificing a bit of accuracy, we can significantly increase the resilience of our system to adversarial examples.
\item We also show empirically that the statistical difference between adversarial and genuine examples is significantly higher when feature selection is applied. This supports the validity of our normalised metrics and the argument that embedded feature selection mitigates evasion attacks, as the detection of adversarial examples is easier when the number of features is reduced.
\end{itemize}

The rest of the paper is organised as follows: In Section 2 we provide the background and relevant related work. In Section 3 we describe our design and testing methodology to evaluate the robustness of feature selection against attacks at test time. In Section 4 we provide a comprehensive experimental evaluation showing the benefits of feature selection to enhance the security of machine learning systems and the trade-off between security and accuracy. Finally, Section 5 discuss the main contributions of the paper and sketch future research avenues. 
\section{Related Work} \label{sec:2}
A comprehensive classification of the different threats against machine learning systems is described in \cite{huang}, providing a taxonomy of attacks according to various criteria. An updated and revised threat model can be found in \cite{papernotTowards}. In this paper we focus on evasion attacks, i.e. those produced at test time, targeting machine learning classifiers.

Szegedy et al. \cite{szegedy} were the first to show the existence of adversarial examples in neural networks and deep learning systems, but as shown in \cite{huang}, previous works in adversarial machine learning analysed evasion attacks in the context of linear classification. Thus, Lowd and Meek \cite{lowd} proposed an algorithm to reverse engineer linear classifiers. Nelson et al. \cite{NelsonEvasion} extended previous work to consider evasion attacks in convex-inducing classifiers. In \cite{biggioEvasion}, Biggio et al. proposed an evasion attack where the adversary aims to find the perturbation that minimises the discriminant function for the adversarial example given the maximum level of perturbation allowed for the attacker. 

A different approach to model the attacker's problem is usually considered in the literature \cite{szegedy,papernotLimitations,carliniTowards} where the attacker aims to find the minimum perturbation that achieves evasion. This can be written as:\footnote{Variants of this attack formulation can be found across different papers in the research literature.}
\begin{equation} \label{eqAttack}
x^* = \argmin_{x'} \ d(x', x) \quad \text{s.t.} \quad F(x') = y_t
\end{equation} where \(d\) is a distance function, $x$ is the sample to be perturbed, $F(x)$ is the predictive function of the classifier and \(y_t\) is the target label. Different distance functions can be considered by the attacker, such as $L1$, $L2$ or $L_\infty$ norms. Thus, the attacker aims to find the minimum distortion that misclassifies the adversarial example as $y_t$. We can model both indiscriminate and targeted attacks depending on the class of errors that the attacker wants to produce. 


Solving the previous optimisation problem can be non-trivial for some machine learning architectures, such as deep learning \cite{larochelle} where the problem is non-convex and non-linear. On the other hand, scalability problems in high-dimensional datasets hinders the practical application of these attack strategies. More scalable and fast attacks have already been proposed in the research literature \cite{goodfellow,kurakinScale,moosavi2016,moosavi2017}, showing that even an approximation to the optimal formulation in (\ref{eqAttack}) is still very effective in successfully evading machine learning systems.

On the defensive side, several mechanisms have already been proposed to either correctly classify or detect adversarial examples. However, Carlini and Wagner \cite{carliniBypassing} have shown that some of the strongest detection techniques can be bypassed by crafting attacks that target them specifically. Amongst the techniques that aim to correctly classify attacks, \emph{adversarial training} has been proven to help mitigate the effect of evasion attacks by augmenting the training data with adversarial examples \cite{szegedy}. Madry et al. \cite{madry} proposed to compute adversarial perturbations at training time to make the learning algorithms more robust to adversarial examples. Kurakin et al. \cite{kurakinScale} used a \emph{single-step} attack to train an \emph{Inception v3} model \cite{szegedy2016} adversarially by linearising the model's loss. Tramer et al. \cite{tramer2018} proposed to augment the training data with adversarial samples transferred from different models, increasing the robustness against black-box attacks.

In this paper we propose a different approach to enhance the robustness of machine learning systems against adversarial examples by reducing the model complexity with feature selection. Bhagoji et al. \cite{bhagoji} proposed to use dimensionality reduction through linear transformation with PCA to enhance the security of machine learning systems. The authors claim that by eliminating components with low variance, the attacker is limited to less optimal perturbations. They also proposed \emph{anti-whitening} as a mechanism to exaggerate the disparity between the variances of the components in PCA. Thus, instead of \emph{cutting off} low variance components, it increases the price of accessing them.

Biggio et al. \cite{biggioSecurity} proposed a framework to evaluate empirically the security of machine learning classifiers at design time. They provide some experimental evidence, using information gain in filter-based feature selection technique. They concluded that feature selection makes models less robust against evasion attacks. The experimental evidence is only provided for $L_0$ norm attacks using binary features. However, the experimental comparison is not fair, as the models with different features counts are compared according to the total level of distortion. As we will show in the next sections, this methodology can produce misleading results. Furthermore, filter-based feature selection often produces less optimal models than wrapper and embedded feature selection techniques. 

Zhang et al. \cite{zhangAdversarial} proposed an adversary-aware feature selection approach for binary classifiers based on an optimisation problem that considers at the same time the generalisation capabilities of the learning algorithm (i.e. the classification accuracy) and the robustness against evasion attacks (measured as the average minimum number of modifications required to evade the classifier). The proposed optimisation problem can be approximately solved using \emph{wrapper-based} feature selection such as forward selection and backward elimination. As in \cite{biggioSecurity}, the experimental evaluation suggests that feature selection makes models less secure, as the attacker is required to modify fewer features to succeed in the attack. However, as in the previous case, the comparison between models with different features is based on the absolute number of modifications, not on the relative perturbation per feature. The proposed methodology does not provide an aggregate measure to assess the security of machine learning classifiers against different types of attacks. On the other hand, the authors use greedy wrapper feature selection methods, which usually produce less optimal results compared to embedded feature selection.

Finally, Wang et al. \cite{wangSparse} devised a game between a regularised (binary) classifier and  an attacker under white-box settings. The game starts with the attacker injecting worst-case perturbations to positive examples. Then, the classifier re-adjusts its parameters on the dataset altered by the attacker. The algorithm is repeated until convergence. The experimental evaluation in \cite{wangSparse} shows that, as the attack strength increases, the accuracy of $L_1$ (Lasso) regularised models degrade more gracefully compared to $L_2$ regularised ones. However, the attacker model proposed is suboptimal, as the attacker selects features to modify at random. There is also no comparison against non-regularised models to assess whether $L_1$ and $L_2$ are more secure against evasion attacks. 
\section{Mitigation of Evasion Attacks with Embedded Feature Selection} \label{sec:3}
Feature selection techniques aim to select subsets of relevant features to reduce model complexity while minimising the loss of information. In this section, we propose an evaluation methodology to assess the security of machine learning classifiers against different types of evasion attacks when feature selection is applied. We provide a mechanism to compare the attack strengths required to craft successful adversarial examples for models with different feature counts by considering the average normalised minimum distortion introduced by the attacker.  

In the rest of this section, we first provide motivation for the use of feature selection to mitigate the effect of evasion attacks, which includes a synthetic example. Then, we briefly review embedded feature selection methods, focusing on Lasso, and the attack strategies used to craft successful adversarial examples with minimum perturbations. Finally, we describe our security evaluation methodology.

\subsection{Motivation}
Adversarial examples are possible in regions of sample space where there is a mismatch between the \emph{ideal}, true model, which we would learn if we could completely characterise the data distribution, and the model learned by the machine learning algorithm on a finite set of training points. Transferability of attacks \cite{papernotTransferability} suggests that these \emph{adversarial regions} are, to some extent, shared across different machine learning models. It is reasonable to hypothesise that the transferability of attacks is due to the lack of data in the adversarial regions, i.e. adversarial examples can be produced in regions where genuine data does not occur naturally (or occurs infrequently), but that are sufficiently \emph{close} to regions where the probability density of data points is high. This makes adversarial examples difficult to detect, especially when the number of features is high. \textit{Adversarial training} aims to inject adversarial examples in the training set to fill in some of these adversarial regions to make the system more robust. 

In this paper, we propose to use a different approach to deal with model complexity that can be complementary to adversarial training. Complex models require more training data to produce more accurate predictions and to reduce the mismatch between the \emph{ideal} model and the learned model.\footnote{Provided that the learning algorithm has enough expressive power, e.g. a linear classifier cannot approximate the true decision boundary in a non-linear problem even if we had an infinite number of training points.} Feature selection is a well-known mechanism to reduce model complexity, eliminating variables that are irrelevant, redundant, or provide little information to solve the learning task. Thus, by reducing the feature set used by the learning algorithm we also reduce model complexity. However, reduction in the number of features implies a trade-off between accuracy and complexity, i.e. performance can decrease when reducing the number of features. However, by reducing model complexity with feature selection, we expect to shrink the regions that can be leveraged by the attacker to craft adversarial examples. 

\begin{figure}
\centering
\includegraphics[width=0.5\textwidth]{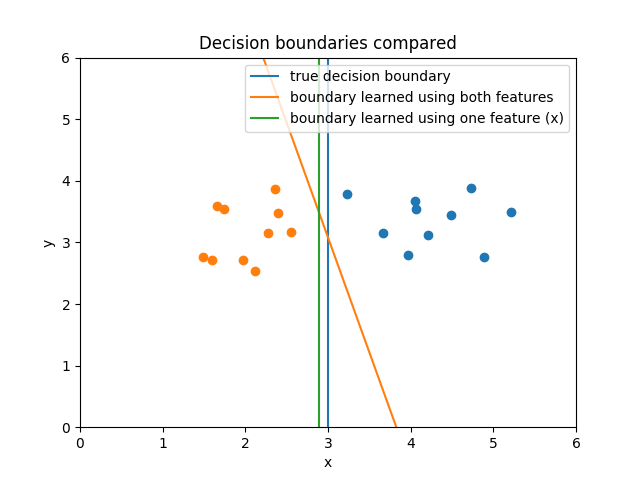}
\caption{Synthetic binary classification dataset where only one of the features is relevant to solve the classification task. Blue line depicts the Bayes optimal decision boundary; green line depicts the decision boundary learned by an SVM considering only the relevant feature; orange line represent the decision boundary of an SVM that learns using the 2 features. It can be appreciated that the mismatch between the true and the learned decision boundary is higher when we use the two features to train the SVM, which increases the opportunity for attackers to craft successful adversarial examples.}
\label{fig:svm_3}
\end{figure}

To illustrate this, we show in Figure \ref{fig:svm_3} a 2-dimensional synthetic example for binary classification where the first feature of class 0 data points was drawn from the Gaussian distribution \(\mathcal{N}(2, 0.25)\), whereas the first feature of class 1 points was drawn from \(\mathcal{N}(4, 0.25)\). The second feature of all data points was drawn from \(\mathcal{N}(3, 0.25)\). Hence, class 0 and class 1 instances only differ in the distribution of the first feature. We generated 10 points for each class and trained two SVM classifiers: one using both features and another using only the first (relevant) feature. As we know the underlying data distribution, we can also characterise the Bayes optimal decision boundary. In Figure \ref{fig:svm_3}, we can observe that if both features were used during training, the difference between the learned and true decision boundaries (i.e. the adversarial region) is large, as the hyperplane fits to the noise of the irrelevant feature. In contrast, training with only the relevant feature yields a decision boundary much closer to the true one. 
 
Typically, adversarial examples leverage some of the most important features for the learning algorithm, allowing small changes in the inputs to produce large changes to the output. However, less important features also need to be modified sometimes to craft optimal attacks. Thus, by eliminating unimportant features, the attacker can no longer manipulate them to change the decision output. As a result, the relative perturbation per feature that the attacker needs to introduce increases, i.e. with fewer features, the attack distance might account for a larger proportion of the maximum possible attack distance. Hence, we can expect adversarial samples to show stronger discrepancy with clean samples in reduced space. This could make it easier for statistical anomaly detection techniques to detect attacks. Our experimental evaluation in Section 4 validates this hypothesis.

\subsection{Embedded Feature Selection}
Feature selection methods are usually grouped into three categories according to how feature selection interacts with the learning algorithm. \emph{Filter methods} do not consider the learning algorithm at all when selecting the features. They usually rely on the mutual information between each feature and the predicted variable. \emph{Wrapper methods} rely on evaluating the performance of classifiers trained on different subsets of features. Although they consider the dependencies between the features, these methods are computationally demanding for large sets of features, as they require the model to be retrained for every feature subset evaluated. Finally, \emph{embedded feature selection} incorporates feature selection into the training process of the learning algorithm.

Lasso \cite{tibshiraniLasso} is a well-known method for the regularisation of linear regression models, where introducing a constraint on the $L1$ norm of a model's parameters (excluding the bias) improves its generalisability and, at the same time, performs variable selection, which is achieved by introducing \emph{sparsity} in the solution, i.e. the coefficients of the less relevant features a zeroed. 

Given a cost function $C(\mathcal{D}_{tr},{\bf w})$ evaluated on a training dataset $\mathcal{D}_{tr}$, the parameters of the model can be learned by solving the optimisation problem:
\begin{equation}
{\bf w}^* \in \min_w C(\mathcal{D}_{tr},{\bf w}) + \lambda \Vert {\bf w} \Vert_1.
\label{eqLasso}
\end{equation} where the regularisation parameter $\lambda$ controls the sparsity of the model, performing feature selection automatically. Thus, larger values of $\lambda$ induce models with less features. 

$L2$ penalty terms are also used as a regularisation technique to prevent overfitting. L2 regularisation does not force model sparsity, so it cannot be applied as feature selection \emph{per se}, but it is possible to perform feature selection by zeroing weights whose magnitudes are below a given threshold. However, if two or more features are highly correlated, then it is reasonable to keep only one of the features to simplify the model complexity. Regardless, $L2$ regularisation will distribute weights somewhat equally among them, which could result in all of these features being removed or kept. In contrast, Lasso ($L1$ regularisation), does not suffer from this problem and automatically selects the best features to keep.

\subsection{Attack Strategy} \label{sec:attack}
We evaluate the robustness of classifiers by crafting minimal attacks, defined in equation (\ref{eqAttack}). A minimal attack is an adversarial sample that barely causes the classifier to misclassify it as the target class and whose distance to the source sample is minimal according to some distance metric. Such attacks allow us to establish the lower bound of a classifier's robustness.

To find the minimal attack for a sample, we use the Jacobian-based Saliency Map Attack (JSMA) \cite{papernotLimitations} to solve
\begin{equation} \label{eq:2}
x^* = \argmax_{x'} \ \ell(f(x'), y) \quad \text{s.t.} \quad d(x',\,x) < \Gamma
\end{equation}
where \(\ell(f(x'), y)\) is the cross entropy loss between the classifier output $f(x)$ and the original class, and \(\Gamma\) is the maximum attack distance. Solving the above optimisation problem yields an adversarial sample that maximises the prediction error under a given distance budget. An approximate solution to (\ref{eqAttack}) can be obtained by performing binary search on \(\Gamma\).\par
Algorithm (\ref{alg:adv}) shows our adaptation of JSMA to solve (\ref{eq:2}) under \(L_1\), \(L_2\) and \(L_{\infty}\) distance constraints. We enforce the attack norm constraints by projecting the perturbation vector \(\delta^*\) onto \(L_p\) space in each iteration of the gradient ascent on the optimisation problem. \(L_2\) (Algorithm \ref{alg:l2_proj}) and \(L_{\infty}\) (Algorithm \ref{alg:linf_proj}) projections are straightforward, whereas the projection onto the \(L_1\) (Algorithm \ref{alg:l1_proj}) space is more involved and requires approximations to compute it efficiently. For this, we use the \(\mathcal{O}(n\log{}n)\) algorithm introduced by Duchi et al.\cite{duchi2008efficient}.\footnote{While they also provided a linear time algorithm in the same paper, we found the \(\mathcal{O}(n\log{}n)\) version to perform 3x times as fast in our implementation.} 

\begin{algorithm}[H]
\caption{Attack algorithm}
\label{alg:adv}
\begin{flushleft}
\textbf{Input}: \(\boldsymbol{X}\): source sample; \(\boldsymbol{Y}\): original class; \(\boldsymbol{Y^*}\): target class; \(l\): cross entropy loss; \(f\): output of network; \(\Gamma\): maximum distance; \(p\): attack norm; \(\alpha\): step size \\
\textbf{Output}: \(\boldsymbol{X^*}\): adversarial sample
\end{flushleft}
\begin{algorithmic}[1]
\State \(\boldsymbol{X^*} \gets \boldsymbol{X}\)
\While{\(f(\boldsymbol{X^*}) \neq \boldsymbol{Y^*}\)}
	\State \(\delta \gets \alpha \cdot \frac{\partial l(f(\boldsymbol{X^*}), \boldsymbol{Y})}{\partial \boldsymbol{X^*}}\)
    \State \(\delta^{\prime} \gets \boldsymbol{X^*} -\boldsymbol{X} + \delta \)
    \State \(\delta^* \gets project_{L_p}(\delta^{\prime}, \Gamma)\)
    \State \(\boldsymbol{X^*} \gets \boldsymbol{X} + \delta^*\)
\EndWhile
\end{algorithmic}
\end{algorithm}

\begin{algorithm}[H]
\caption{\(project_{L_1}\) \cite{duchi2008efficient}}
\label{alg:l1_proj}
\begin{flushleft}
\textbf{Input}: \(\boldsymbol{v} \in \mathbb{R}^n\) : vector to be projected; \(d \in \mathbb{R} > 0\): maximum \( L_1\)-norm of the projected vector \\
\textbf{Output}: \(\boldsymbol{w} \in \mathbb{R}^n\): projected vector which solves \(\min_{\boldsymbol{w}^{\prime}} \Vert {\boldsymbol{w}^{\prime}} - \boldsymbol{v} \Vert_2^2\) s.t. \(\Vert {\boldsymbol{w}^{\prime}} \Vert_1 \leq d\)
\end{flushleft}
\begin{algorithmic}[1]
\State Define \(\boldsymbol{u}\) where \(u_i = \vert v_i \vert\)
\State \(\boldsymbol{\mu} \gets \boldsymbol{u}\) sorted in descending order
\State \(\rho \gets \max \left\{j \in \left\{ 1,...,n \right\}: \mu_j - \frac{1}{j} \left( \sum_{r=1}^j \mu_r - d \right) > 0\right\}\)
\State \(\theta \gets \frac{1}{\rho} \left( \sum_{i=1}^{\rho} \mu_i - d \right)\)
\State Define \(\boldsymbol{z}\) where \(z_i = \max \left\{ u_i - \theta, 0\right\}\)
\State \(\boldsymbol{w} \gets sign(\boldsymbol{v}) \star \boldsymbol{z}\)
\end{algorithmic}
\end{algorithm}

\begin{algorithm}[H]
\caption{\(project_{L_2}\)}
\label{alg:l2_proj}
\begin{flushleft}
\textbf{Input}: \(\boldsymbol{v} \in \mathbb{R}^n\) : vector to be projected; \(d \in \mathbb{R} > 0\): maximum \( L_2\)-norm of the projected vector \\
\textbf{Output}: \(\boldsymbol{w} \in \mathbb{R}^n\): projected vector which solves \(\min_{\boldsymbol{w}^{\prime}} \Vert {\boldsymbol{w}^{\prime}} - \boldsymbol{v} \Vert_2^2\) s.t. \(\Vert {\boldsymbol{w}^{\prime}} \Vert_2 \leq d\)
\end{flushleft}
\begin{algorithmic}[1]
\State \(n \gets \Vert \boldsymbol{v} \Vert_2\)
\State \(\boldsymbol{w} \gets \frac{\boldsymbol{v} \cdot d}{n} \)
\end{algorithmic}
\end{algorithm}

\begin{algorithm}[H]
\caption{\(project_{L_{\infty}}\)}
\label{alg:linf_proj}
\begin{flushleft}
\textbf{Input}: \(\boldsymbol{v} \in \mathbb{R}^n\) : vector to be projected; \(d \in \mathbb{R} > 0\): maximum \( L_{\infty}\)-norm of the projected vector \\
\textbf{Output}: \(\boldsymbol{w} \in \mathbb{R}^n\): projected vector which solves \(\min_{\boldsymbol{w}^{\prime}} \Vert {\boldsymbol{w}^{\prime}} - \boldsymbol{v} \Vert_2^2\) s.t. \(\Vert {\boldsymbol{w}^{\prime}} \Vert_{\infty} \leq d\)
\end{flushleft}
\begin{algorithmic}[1]
\State Define \(\boldsymbol{w}\) where \(w_i = \min \left\{ \max \left\{w_i, -d \right\} , d \right\}\)
\end{algorithmic}
\end{algorithm}

\subsection{Security Evaluation}
\subsubsection{Distance Metric}
Related works on feature selection in adversarial settings \cite{biggioSecurity,zhangAdversarial} use the same metric to compare the robustness of models with different input dimensions. We argue that this is not appropriate because applying feature selection results in a different classification problem from the original one, and the same perturbation applied to the input of a smaller model means a bigger proportional change. This could mean that adversarial examples are easier to detect under reduced feature sets. To account for the differences in input dimensions, we argue that it is necessary to normalise the minimal attack distance. This leads to different results from those reported in the related works.  \par
Simon-Gabriel et al.\cite{simonAdversarial} describe a way to scale the attack strength according to the attack norm and input dimension. To preserve the average \emph{signal-to-noise} ratio \(\Vert x \Vert_2 / \Vert \delta \Vert_2\), they suggest to compute the attack strength used with \(L_p\)-attacks as 
\[\Gamma_p = c \ f^{1/p}\]
where \(c\) is a dimension-independent constant and \(f\) is the input dimension.
The scaling is useful for normalising attack strengths across different models for a fixed value of \(c\). This allows metrics such as evasion rate to be measured against the fixed normalised attack strength. However, if we use minimal adversarial samples, the attack strength is variable. Hence, we propose to divide the minimal attack distance by the normalising factor \(f^{1/p}\):
\begin{equation}
\Gamma_p = \frac{\Gamma_{min}}{f^{1/p}} \in \left[ 0,1 \right].
\label{eq:distort_ratio}
\end{equation}
Assuming all feature values are between 0 and 1, the normalised attack distance denotes the ratio between the distortion strength and the maximum possible distortion strength. This metric captures the percentage of distortion as decimal and can be used to compare models with different input dimensions intuitively. We theorise that it is correlated with statistical deviation from clean samples and prove this empirically in Subsection \ref{ssec:mmd}. \par
To measure the robustness of a model against attacks constrained by \(L_p\) norm, we craft minimal \(L_p\) attacks with Algorithm (\ref{alg:adv}) from a set of samples and compute the average of the normalised attack distances across these samples.

\subsubsection{Security Score}
As the defender cannot anticipate the type of attack an adversary will use, the distortion percentage for a specific attack norm is by itself insufficient to measure a model's general robustness against attacks. We propose to use an aggregated security \(s \in \left[ 0,1 \right]\) by averaging over the normalised attack distances for all attack norms used in the security evaluation:
\begin{equation}
s = \frac{1}{\vert P \vert} \sum_{p \in P} \frac{\Gamma_p}{f^{\slfrac{1}{p}}}
\label{eq:sec_score}
\end{equation}
where \(P\) is the set of all attack norms considered.
Comparing \(s\) against the prediction accuracy allows us to study the trade-off between security and accuracy as we vary the aggressiveness of feature selection. 

\subsubsection{Statistical Analysis of Attacks}
To evaluate how feature selection affects the detectability of adversarial samples, similar to \cite{grosse2017} , we use \textit{Maximum Mean Discrepancy} (MMD) \cite{fortet1953convergence} to measure the statistical distance between adversarial and clean samples. Given two sets of samples drawn from different data distributions, \(X = \left\{ x_1,...,x_n \right\} \sim p\) and \(Z = \left\{ z_1,...,z_n \right\} \sim q\), the MMD is defined as
\[D(X, Z, \mathcal{F}) = \sup_{f \in \mathcal{F}} \mathbf{E}_p \left[f(x)\right] - \mathbf{E}_q \left[f(z)\right]\]
where \(\mathcal{F}\) is a class of functions and \(\mathbf{E}\) is the expectation. \(D(X, Z, \mathcal{F}) = 0\) iff \(p = q\). \par
Its unbiased empirical estimate \cite{gretton2007kernel} can be computed as 
\begin{multline}
\widehat{D}(X, Y) = 
\frac{1}{n^2} \sum_{i=1}^n \sum_{j=1}^n K(x_i, x_j) - 
\frac{2}{nm} \sum_{i=1}^n \sum_{j=1}^m K(x_i, z_j) \\ +
\frac{1}{m^2} \sum_{i=1}^m \sum_{j=1}^m K(z_i, z_j)
\label{eq:mmd_est}
\end{multline}
where \(K\) is a kernel function. \par
Grosse et al.\cite{grosse2017} have shown that the MMD between adversarial and clean samples is significantly higher than the MMD between two sets of clean samples, thus proving the effectiveness of this metric in detecting attacks. \par
While they used a Gaussian kernel in their experiments, we opt to use the normalised linear kernel which does not make any assumptions about the underlying distributions and mitigates discrepancies caused by differences in input dimension:
\[K(x, z) = \frac{x^{\intercal}z}{\Vert x \Vert_1 \Vert z \Vert_1}.\]
Our experimental evaluation in Section \ref{sec:4} shows that the MDD of adversarial examples increases as we reduce the number of features.

\subsection{Feature Representation vs Feature Selection}
It is important to distinguish between feature representation and feature selection, especially when designing machine learning systems with security considerations. In the first case we refer to the set of features considered in the design of the system, regardless of the importance of the features. In contrast, once this set of features is defined, feature selection aims to select the relevant features based on the data available for training the learning algorithm. 

Feature representation not only affects the performance of the system, but it can also have a significant impact on the robustness of the system against evasion attacks, as attackers can evade the machine learning system by using features that are not considered during the design of the system. For instance, in spam filtering applications, attackers can misspell bad words (e.g. by substituting \emph{profit} with \emph{prof1t}), so that if there are no mechanisms in place to deal with such obfuscation techniques, the attacker can easily evade detection, ignoring other possible vulnerabilities of learning algorithm. 

In feature selection, based on the training data available, we remove features that are not important or redundant for the learning task. Although the attacker can leverage the removed features to try to evade the system, if the training data representation is reasonable, possibly the use of these features does not bring important value for the attacker, for example in terms of functionality, to fulfil her goals. 

In some security applications such as malware or anomaly detection, other feature selection techniques can be considered to prevent the attacker from evading the system. In these cases, the attacker typically only aims to produce one type of error in the system, i.e. to craft malicious examples that are misclassified as benign. Therefore, a possible approach is to perform asymmetrical feature selection by removing only features that are indicative of benign behaviour. 

\section{Experiments} \label{sec:4}
In this section we show our experimental evaluation on the security of classifiers trained using embedded feature selection through $L1$ regularisation. We show that by reducing the number of features used by the learning algorithm, we increase the security, as the attacker is required to introduce higher relative distortion in the adversarial examples to succeed. We also show the natural trade-off between security and accuracy in these settings. We observe that some predictive accuracy can be sacrificed to significantly increase a model's security. Finally, we show that adversarial examples crafted with the minimal attack strategy described in Section \ref{sec:attack} statistically differ more strongly from the genuine examples when number of features is reduced, proving that it is easier to detect adversarial examples for reduced feature subsets. This also supports the use of our methodology that considers the relative perturbation added to the adversarial examples rather than the absolute values, as proposed in previous related works.

In our experiments we analyse the security of logistic regression classifiers on 3 real binary classification datasets:
\begin{itemize}
\item \textbf{MNIST} \cite{lecun} is a well-known image dataset containing handwritten digits labelled from 0 to 9. Each data sample is a gray scale $28 \times 28$ image with 784 features, each representing a pixel. We focus on the subset consisting of handwritten 7s and 9s, as they have visually similar components. We assign class 1 to the former and class 0 to the latter. There are 7,293 class 1 and 6,958 class 0 samples in total.
\item \textbf{Ransomware}: This dataset contains data points with 232 features, which is a subset of full feature set in \cite{sgandurra2016automated}. The features represent the frequencies of API invocations obtained during the dynamic analysis of ransomware and benign applications. It includes 828 ransomware samples (class 1) and 942 samples of benign software (class 0).
\item \textbf{PDF Malware}: This is a subset of the PDF Malware dataset used in \cite{biggioEvasion} with 5,993 positive and 5,951 negative samples. There are 114 features representing the occurrence of specific keywords relevant to detect malicious PDF files. We have normalised all the features to the interval $[0, 1]$.
\end{itemize}

In our experimental evaluation, we employ 10-fold cross validation, using 9 folds for training and 1 for crafting the adversarial examples. \par
All models are trained using standard gradient descent on the sigmoid cross entropy loss function, implemented using \emph{Tensorflow}. \par
Different input dimensions are obtained by an automated process that searches for suitable multipliers for the regularisation term. It should be noted that Tensorflow's implementation of gradient descent optimiser does not yield truly sparse weights with $L1$ regularisation, as weights end up with very small magnitudes instead of being zeroed. To get sparse solutions, we specify a small threshold for the weight magnitude under which weights are zeroed. We use a threshold of 0.01 for models trained on MNIST and Ransomware, and 0.1 in the case of PDF Malware.

\subsection{Attack Resistance}
\begin{figure}
\centering
\includegraphics[width=0.4\textwidth]{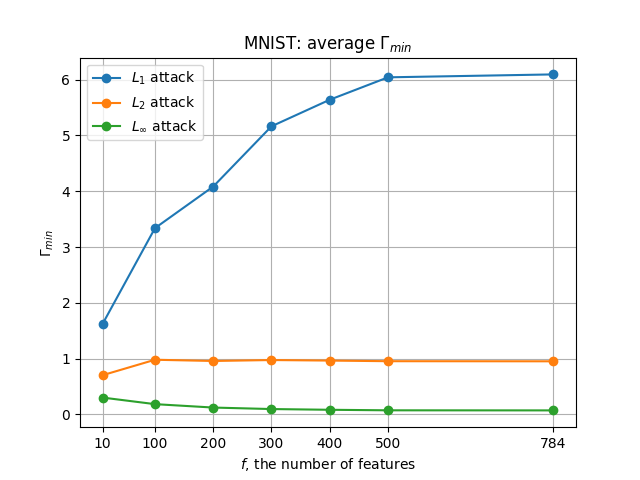}
\caption{Averaged unnormalised minimal attack distance for MNIST as a function of the number of features for \(L_1\), \(L_2\), and \(L_{\infty}\) attacks.}
\label{fig:mnist_atk_unnorm}
\end{figure}

In our first experiment we evaluate the robustness of regularised logistic regression classifiers against adversarial examples crafted with the attack described in Section \ref{sec:attack}. Following a similar presentation to those used in \cite{biggioSecurity,zhangAdversarial}, Figure \ref{fig:mnist_atk_unnorm} shows the average unnormalised minimal attack distances for \(L_1\), \(L_2\) and \(L_{\infty}\) attacks targeting logistic regression classifiers on MNIST with varying levels of regularisation. In the case of \(L_1\) attacks, \(\Gamma_{min}\) increases significantly and monotonically with the number of features, \(f\), with \(\Gamma_{min}\) for the full model being three times as high as \(\Gamma_{min}\) for the smallest model with only 10 features. For \(L_2\) attacks, the average \(\Gamma_{min}\) is stable across all but the smallest model, where the overall unnormalised perturbation is smaller. In contrast to the results for \(L_1\) and \(L_2\) attacks, embedded feature selection with Lasso decreases the minimal distance for \(L_{\infty}\) attacks as \(f\) decreases . \par

\begin{figure}
\centering
\includegraphics[width=0.4\textwidth]{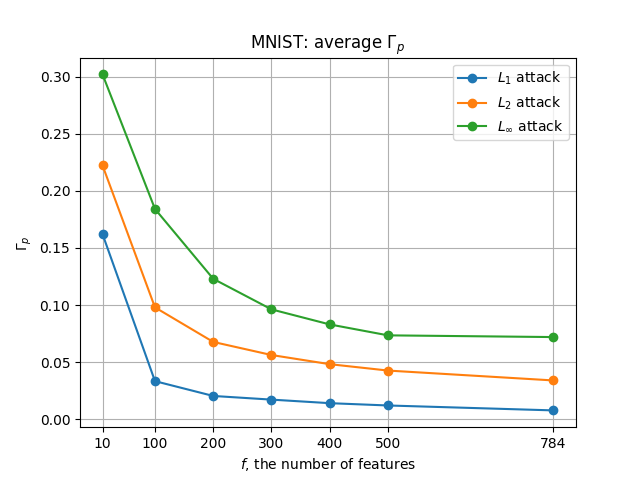}
\caption{Averaged normalised minimal attack distance for MNIST as a function of the number of features selected for \(L_1\), \(L_2\), and \(L_{\infty}\) attacks.}
\label{fig:mnist_atk_norm}
\end{figure}

Based on the results in Figure \ref{fig:mnist_atk_unnorm} alone, it is easy to jump to the conclusion that embedded feature selection can have a negative impact on the overall security of classifiers, as the overall perturbation needed to craft successful \(L_1\) and \(L_2\) attack samples is smaller for reduced feature sets. However, it is no coincidence that the opposite effect is observed for \(L_{\infty}\) attacks, where the distance is naturally normalised according to (\ref{eq:distort_ratio}). \par
Figure \ref{fig:mnist_atk_norm} shows the security analysis of classifiers with different numbers of features when the distance values are normalised. We can observe that feature selection increases the attacker's proportional effort significantly for the 3 attacks as \(f\) decreases. Comparing the model with all the features (784) to the model with only 10 features, the normalised distortion increases 4 times for \(L_1\) attacks, 7 times for \(L_2\) attacks, and 10 times for \(L_\infty\) attacks. \par

\begin{figure}
\centering
\includegraphics[width=0.4\textwidth]{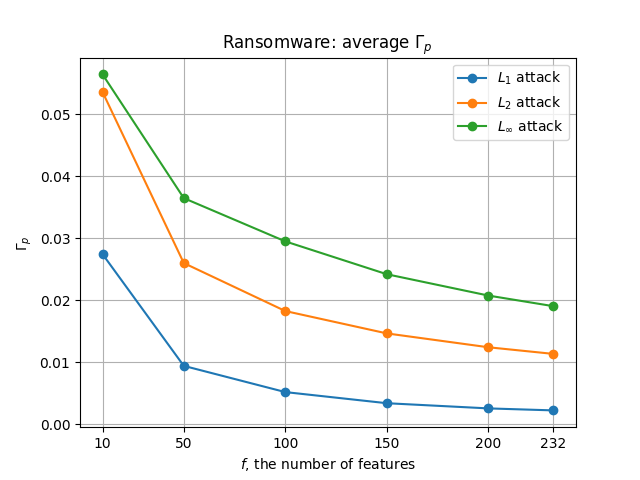}
\caption{Averaged normalised minimal attack distance to evade an $L1$ regularised logistic regression classifier with different number of features in Ransomware dataset.}
\label{fig:ransom_atk_norm}
\end{figure}

\begin{figure}
\centering
\includegraphics[width=0.4\textwidth]{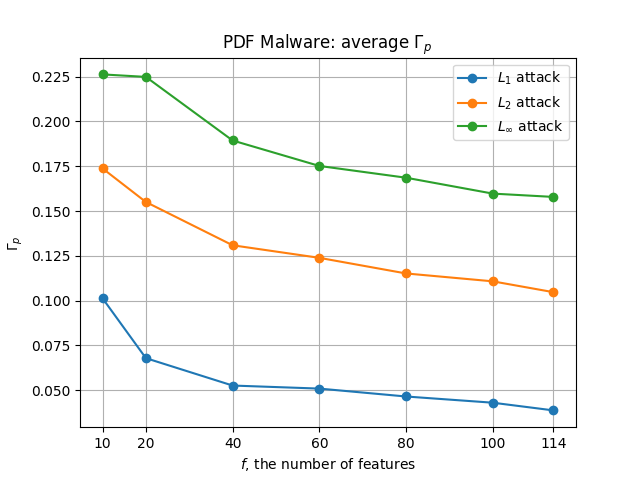}
\caption{Averaged normalised minimal attack distance to evade an $L1$ regularised logistic regression classifier with different number of features in PDF Malware dataset.}
\label{fig:pdf_atk_norm}
\end{figure}

Similar results are shown in Figures (\ref{fig:ransom_atk_norm}) and (\ref{fig:pdf_atk_norm}) for the Ransomware and PDF Malware datasets respectively. In all cases the normalised minimal attack distance decreases with the number of features, showing that the attacker has to increase the relative effort per feature to successfully evade the logistic regression classifier.

\subsection{Accuracy vs. Security Trade-off}
Reduction in the number of features can also lead to a decrease in the performance of a machine learning system, especially with strongly reduced feature sets. Thus, in view of the previous results, we can expect a natural trade-off between security and accuracy. We will now characterise this tradeoff in our examples. It is important to be able to do so in order to choose appropriately between sacrificing accuracy and increasing robustness. In our second experiment we compute the security score \(s\) given in (\ref{eq:sec_score}) for each model. As in the previous experiment we consider \(L_1\), \(L_2\), and \(L_{\infty}\) attacks.

\begin{figure}
\centering
\includegraphics[width=0.4\textwidth]{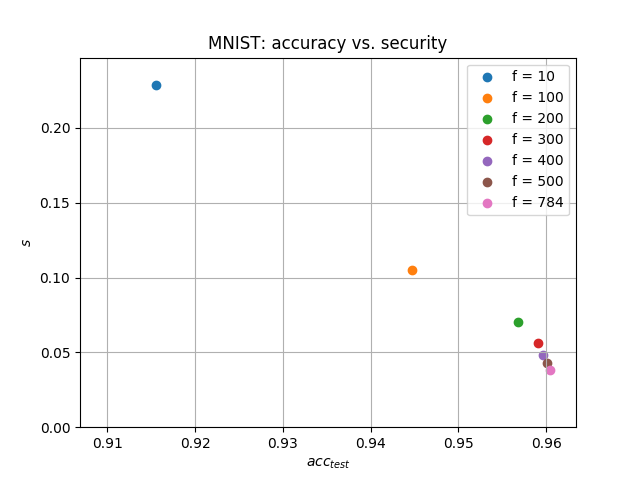}
\caption{Trade-off between prediction accuracy and security score on MNIST dataset.}
\label{fig:mnist_tradeoff}
\end{figure}

Figure \ref{fig:mnist_tradeoff} plots \(s\) against the prediction accuracy of logistic regression classifiers with different number of features for the MNIST dataset. As expected, the accuracy decreases and the security increases with the feature count. The model with 10 features is 5 times more secure than the model with all the features with a reduction in performance of about $4.5\%$. With 100 features, the model is approximately 2 times more secure with a reduction in performance of about $1.5\%$ compared to the complete model. \par

\begin{figure}
\centering
\includegraphics[width=0.4\textwidth]{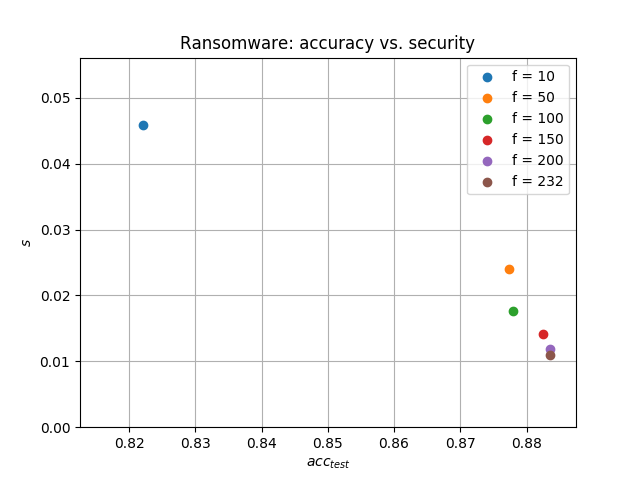}
\caption{Trade-off between prediction accuracy and security score on Ransomware dataset.}
\label{fig:ransomware_tradeoff}
\end{figure}

Figure \ref{fig:ransomware_tradeoff} shows similar results for the Ransomware dataset, where the model with 10 features is approximately 3.5 times more secure with a reduction in performance of about $6\%$. We can also appreciate that models with 50 features doubles the security of the system with a very small loss in performance. \par 

\begin{figure}
\centering
\includegraphics[width=0.4\textwidth]{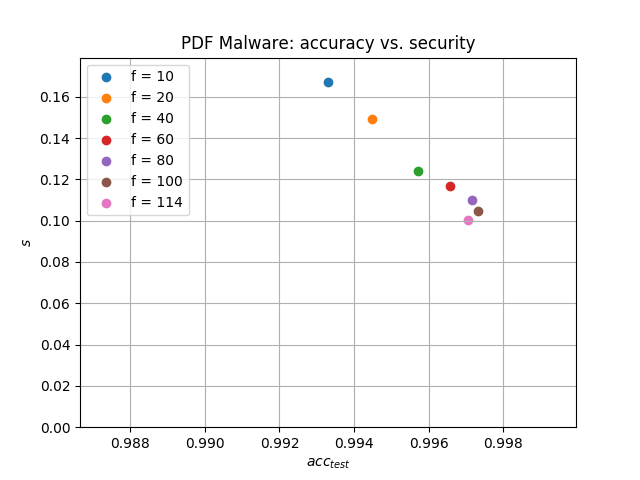}
\caption{Trade-off between prediction accuracy and security score on PDF Malware dataset.}
\label{fig:pdfmalware_tradeoff}
\end{figure}

Finally, the results for the PDF Malware dataset in Figure \ref{fig:pdfmalware_tradeoff} shows that with a low loss in accuracy, we can significantly increase the security of the machine learning classifier. For example, a model with 10 features increases robustness by $0.7$ times with a \(0.3\%\) drop in accuracy.

These representations show the trade-off between accuracy and security, which can be helpful for the design of machine learning systems with security considerations. They allow us to compare models with different number of features, which helps us select operations points according to different security and performance criteria. The results suggest that for each classification task, small sacrifices in accuracy can be made for large increases in robustness. Similar analyses can be performed by looking at the security as a function of the false positive or false negative rates of the machine learning classifiers.

\begin{figure*}
\centering
\includegraphics[width=1\textwidth]{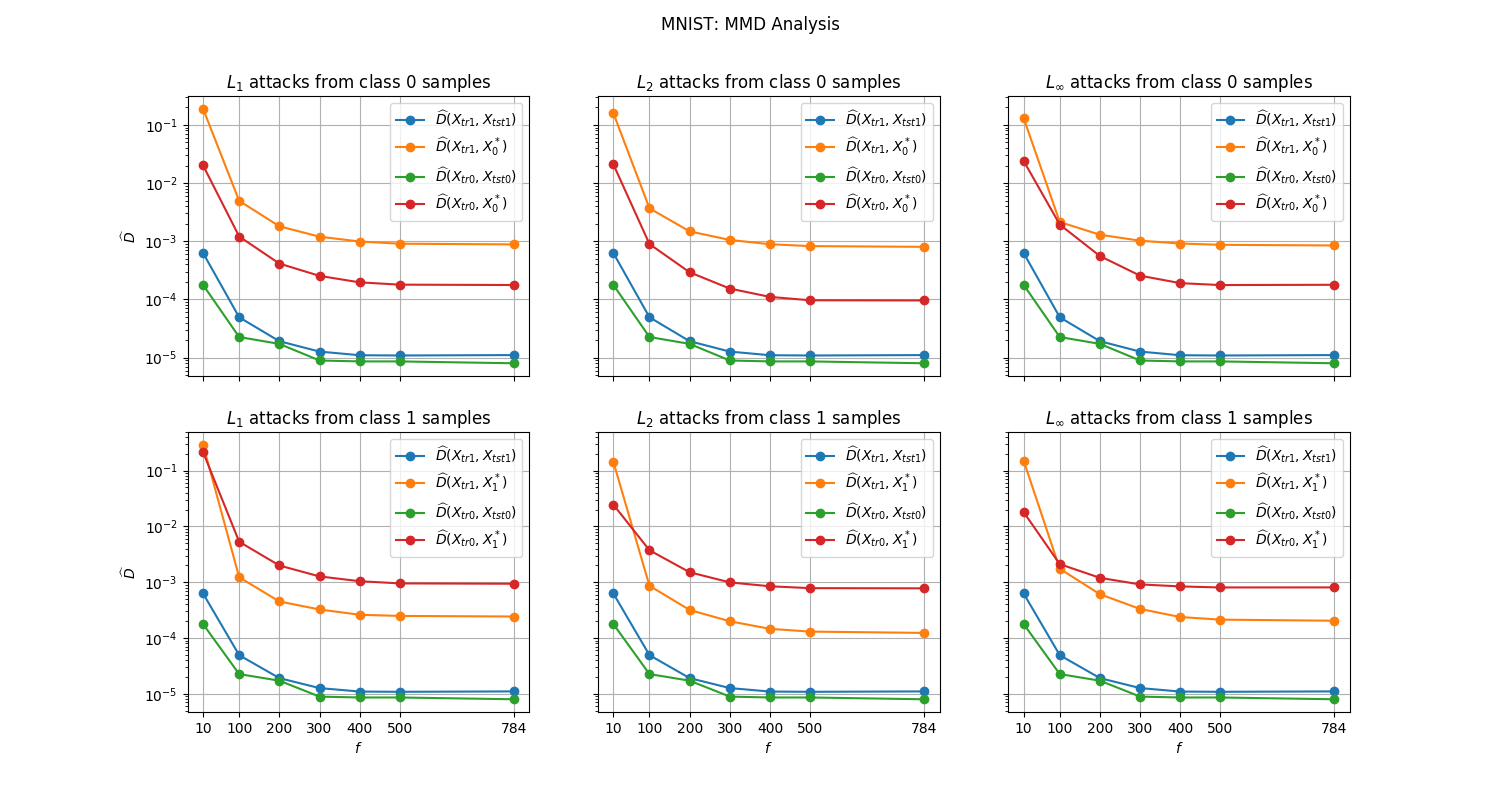}
\caption{MMD estimates for the MNIST dataset and minimal attacks with different feature counts. The different plots are separated by the attack norm and the source class of attacks. \(X_{tra}\), \(X_{tsta}\) and \(X^*_a\) denote the set of class \(a\) training, test and adversarial samples respectively. Note that the y-axis is in logarithmic scale, as some values are orders of magnitudes higher than others. A linear scale version of the plots can be found in Appendix \ref{sec:appendix}, highlighting the differences in the results.}
\label{fig:mnist_mmd_log}
\end{figure*}

\subsection{Statistical Analysis of Adversarial Samples} \label{ssec:mmd}
In the final experiment we provide some empirical support for our evaluation methodology which considers the normalised distortion as opposed to the unnormalised distortion metric used in previous related work \cite{biggioSecurity,zhangAdversarial}. We show that as the relative effort to craft minimal adversarial examples increases when the number of features decreases, the statistical difference between adversarial and genuine samples grows as well.

For the 3 datasets and the 3 attacks considered before, we compute the MMD estimates \(\widehat{D}\) (\ref{eq:mmd_est}) between clean training samples and test samples of the same class to serve as baselines. To determine the statistical detectability of attacks, we compute \( \widehat{D}\) between the adversarial samples and clean training samples, separated by class. \par

For the 3 datasets and the 3 attacks considered before, we have computed the MMD estimates \(\widehat{D}\) (\ref{eq:mmd_est}) between clean training samples and test samples of the same class to serve as baselines. To determine the statistical detectability of attacks, we have computed \( \widehat{D}\) between the adversarial samples and clean training samples, separated by class. \par

Figure \ref{fig:mnist_mmd_log} shows the average MMD estimates in logarithmic scale for the MNIST dataset as a function of the number of features kept. We can observe that for all input dimensions, the \(\widehat{D}\) values between training and adversarial samples are 1 to 2 orders of magnitudes higher than the baseline measurements, regardless of the attack norm and source class. As the number of features drops, all MMD measurements grow at roughly the same exponential rate. However, as the ratio of \(\widehat{D}\) between training and adversarial samples to the baseline values is maintained as \(f\) decreases, the gap between the former and latter is in fact exponentially larger with smaller models. In other words, the statistical difference between genuine and adversarial examples grows as the number of features decreases.

In Appendix \ref{sec:appendix} we provide a different view of the same result, showing MMD using a linear scale, which illustrates better the widening of the gap. We additionally show in Appendix \ref{sec:appendix} the corresponding experimental results for Ransomware and PDF Malware datasets. The results and trends are similar to the those for MNIST, so we omitted them in this section for the sake of brevity. \par

The results on \(L_1\) and \(L_2\) attacks show that despite the fact that the raw minimal attack distance decreases with \(f\), minimal attacks exhibit stronger statistical divergence from clean attacks in lower dimensions. This proves that unnormalised distance metrics are inadequate for assessing the robustness of classifiers. In contrast, our normalised distance metrics show similar trends to the statistical divergence of minimal adversarial samples, as smaller models require exponentially stronger proportional perturbation and at the same time, make attacks exponentially easier to detect. Hence, our normalised distance metrics are more representative of the attacker's cost.  \par

As adversarial examples are easier to detect with reduced feature sets, if appropriate detection mechanisms are in place, such as the ones proposed in \cite{grosse2017} which rely on the use of MMD for detection of adversarial exapmles, we can significantly enhance the security of the machine learning system against evasion attacks.

\section{Conclusion} \label{sec:5}
The existence of adversarial examples can hinder the penetration of machine learning in application domains where the security and safety of the system play a critical role. Defending against this threat is challenging as adversarial examples are difficult to classify or detect. Adversarial examples are produced in regions where natural data points are scarce and where the learned model differs from the optimal model we could theoretically learn with knowledge of the underlying data distribution.

In this paper we have shown that embedded feature selection helps to mitigate evasion attacks, reducing the regions that can be leveraged by the attacker to craft successful adversarial examples. We have also shown empirically that the statistical properties of adversarial examples differ significantly from the genuine ones when using a reduced set of features. This can facilitate their detection and further mitigate the effect of evasion attacks.

We have also provided a design methodology to assess the security of machine learning classifiers against adversarial examples under the effect of embedded feature selection. In contrast to previous related work where the security is measured according to the absolute minimum perturbation required to evade the classifier, we propose to use the relative distortion per feature, which allows us to compare models with different features in a fairer way. Under this metric, embedded feature selection produces more secure machine learning classifiers. The statistical analysis of adversarial examples with MMD supports the validity of our approach. Based on this metric, we have also proposed a more general security score that relies on the aggregation of the normalised minimum perturbation for different attack strategies, where the perturbation is measured with different distance metrics. By using this security score we have shown that there exists a natural trade-off between accuracy and security so that, in some cases, small reductions in the performance of the algorithm can notably increase its security. Thus, our security metric can help design and deploy more secure machine learning systems.

It is also important to distinguish between feature representation and feature selection and understand their security implications. Good feature representations can help to reduce the chances of evasion by the attacker through obfuscation or modification of features that are not considered in the design of the system. On the other hand, feature selection aims to remove features that are not relevant or redundant, given the available data to train the model. Thus, even if the attacker uses the removed features to craft the adversarial example, it is difficult to evade the system, as the use of these features is expected to bring little value or functionality.

Future research will include the analysis of different feature selection schemes including filter and wrapper-methods, as well as the development of asymmetrical feature selection techniques to enhance the security of machine learning in anomaly detection and related security applications.

\begin{figure*}
\centering
\includegraphics[width=1\textwidth]{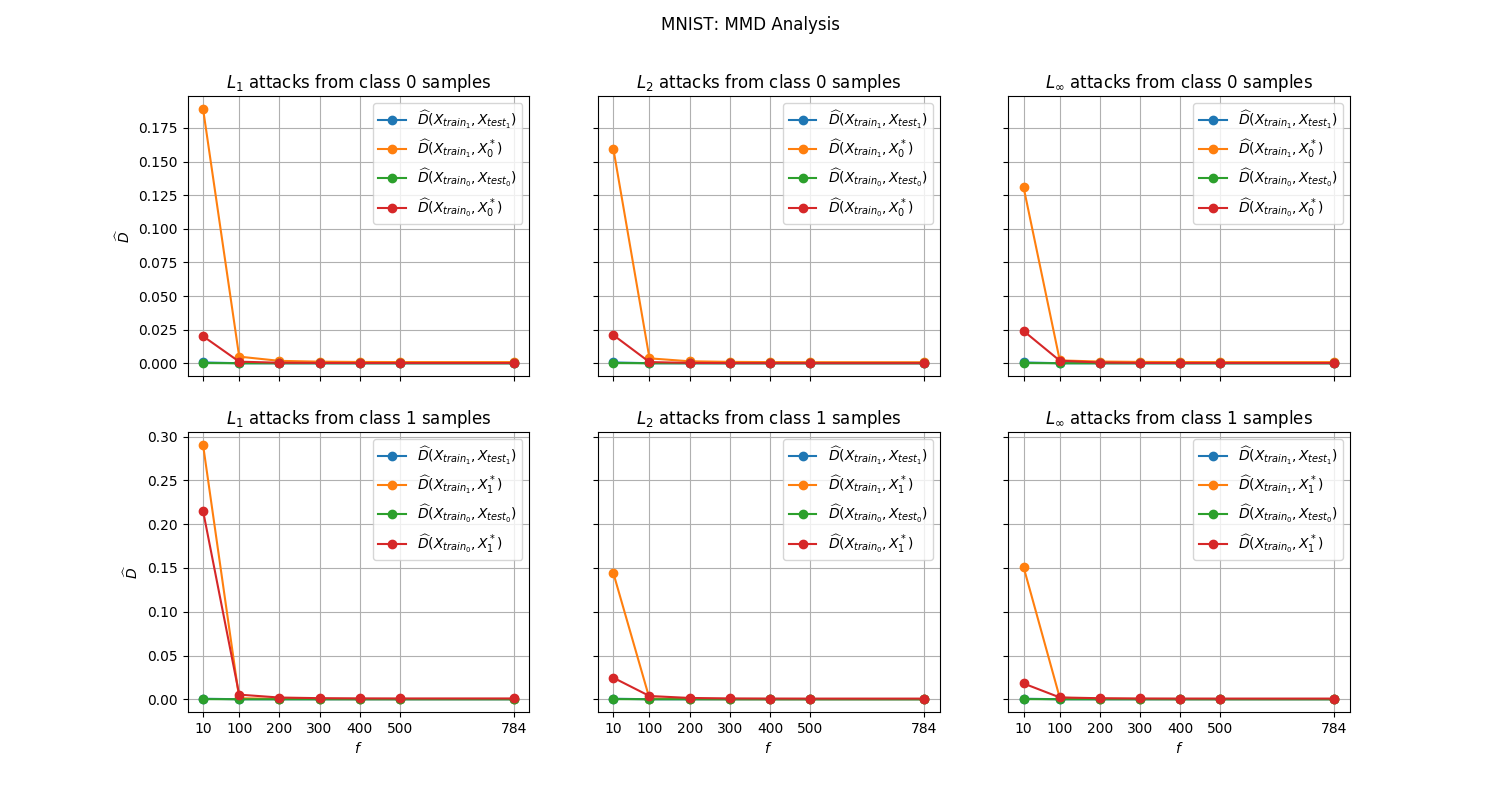}
\caption{MMD estimates for the MNIST dataset and minimal attacks with different feature counts.}
\label{fig:mnist_mmd_lin}
\end{figure*}

\begin{figure*}
\centering
\includegraphics[width=1\textwidth]{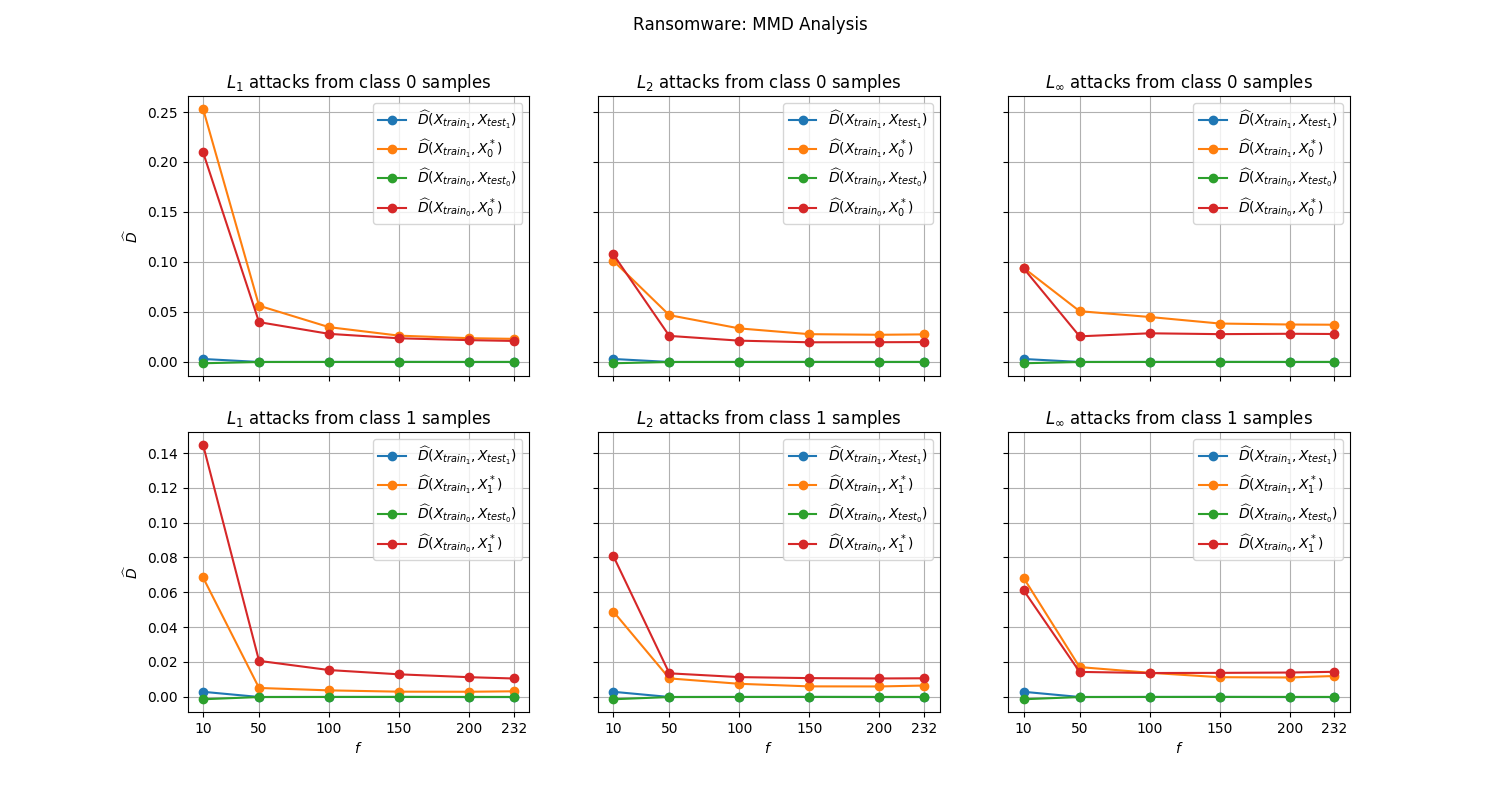}
\caption{MMD estimates for the Ransomware dataset and minimal attacks with different feature counts.}
\label{fig:ransom_mmd_lin}
\end{figure*}

\begin{figure*}
\centering
\includegraphics[width=1\textwidth]{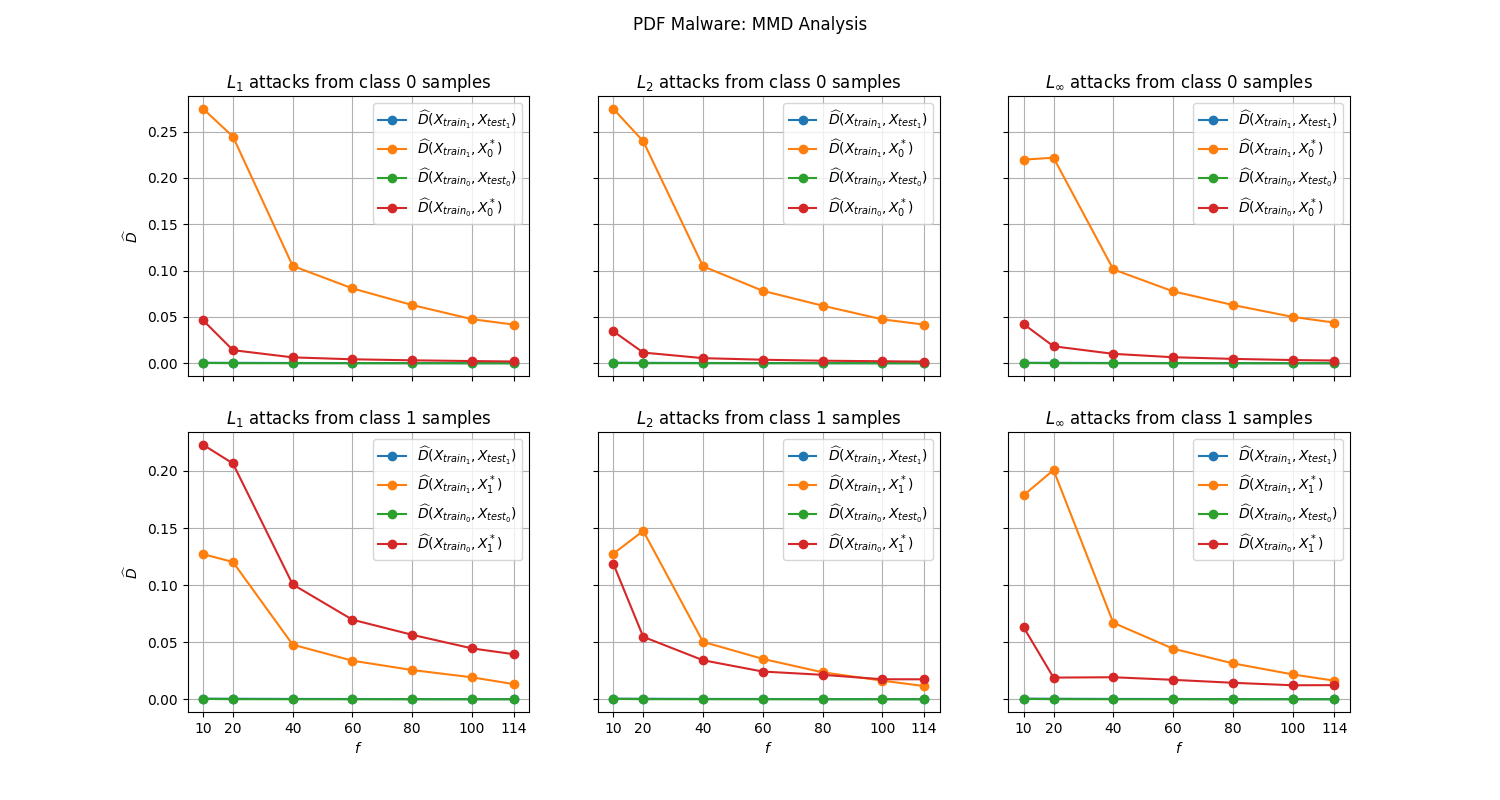}
\caption{MMD estimates for the PDF Malware dataset and minimal attacks with different feature counts.}
\label{fig:pdf_mmd_lin}
\end{figure*}
\bibliographystyle{ACM-Reference-Format}
\bibliography{biblio}

\begin{appendix}
\section{Experimental Analysis of MMD} \label{sec:appendix}
Similar to Figure \ref{fig:mnist_mmd_log}, \ref{fig:mnist_mmd_lin} we show in Figure the MMD analysis with the MNIST dataset on a linear scale, where the significant differences between the properties of the genuine and the adversarial examples are prominent.

Figures \ref{fig:ransom_mmd_lin} and \ref{fig:pdf_mmd_lin} show the same analysis for Ransomware and PDF Malware datasets respectively. The results in these two datasets are similar to those of MNIST.

%
%
\end{appendix}

\end{document}